\documentclass[a4paper]{jpconf}
\usepackage{graphicx}
\usepackage{amssymb}

\begin{document}

\title{Gauge-Higgs unification on the brane}
\author{Gonzalo A Palma$^{1,2}$}
\address{$^1$ Department of Applied Mathematics and Theoretical
Physics, Centre for Mathematical Sciences, University of Cambridge,
Wilberforce Road, Cambridge CB3 0WA, United Kingdom.}

\address{$^2$ D\'epartement de Physique Th\'eorique, Universit\'e de
Gen\`eve, 24 quai Ernest Ansermet, 1211 Gen\`eve 4, Switzerland.}

\ead{G.A.Palma@damtp.cam.ac.uk}

\begin{abstract}
From the quantum field theory point of view, matter and gauge fields
are generally expected to be localised around branes or topological
defects occurring in extra dimensions. Here I discuss a simple
scenario where, by starting with a five dimensional $SU(3)$ gauge theory,
we end up with several 4-D parallel branes with localised ``chiral''
fermions and gauge fields to them. I will show that it is possible to
reproduce the electroweak model confined to a single brane, allowing a
simple and geometrical approach to the fermion hierarchy problem. Some
nice results of this construction are: Gauge and Higgs fields are
unified at the 5-D level; and new particles are predicted: a
left-handed neutrino of zero hypercharge, and a massive vector field
coupling together the new neutrino to other left-handed leptons.
\end{abstract}

\section{Introduction} \label{S1}

In this article I review the talk ``Confining the
electroweak model to a brane'' given during the workshop 
``The Quest for Unification: Theory Confronts Experiment", 
Corfu 2005. I will explain
how, by starting with an $SU(3)$ gauge theory defined in a 5-D
spacetime, we can end up in the presence of a configuration consisting
of several 4-D parallel branes, one of them containing a
copy of the electroweak model. Many of the results discussed here can
be found in ref. \cite{Palma}. I have decided to change the title
of this review in order to emphasise a result which I consider
rather elegant: Some of the components of the original
five-dimensional gauge field, which end up localised to the
``electroweak brane'', can be identified as the Higgs doublet of the
standard model.

To my knowledge, the idea of using extra dimensions to reproduce the
Weinberg-Salam model and unify gauge and Higgs fields into a single
---more fundamental--- field, was first proposed in refs. \cite{Fairlie}
and \cite{Manton} by Fairlie and Manton, respectively. Ever since, there
have been various attempts to find concrete examples of gauge-Higgs
unification \cite{G-H 1}-\cite{G-H 8}.
It has been difficult, however, to
reconcile both fields under the same group representation and have them
coupled to chiral fermions in the appropriate way.
The approach followed here is different from previous works; I first focus in
obtaining the correct chiral structure for leptons and quarks of the
electroweak model, and then verify that the bosonic degrees
coupled to these fermions can be identified with the required Higgs 
doublet and gauge fields. This is done by picking up 
fermions with appropriate $SU(2)\times U(1)$ charges 
from specific $SU(3)$-representations,
and then localise them to a 4-D brane.

\begin{figure}[t] 
\begin{center}
\includegraphics[width=0.77\textwidth]{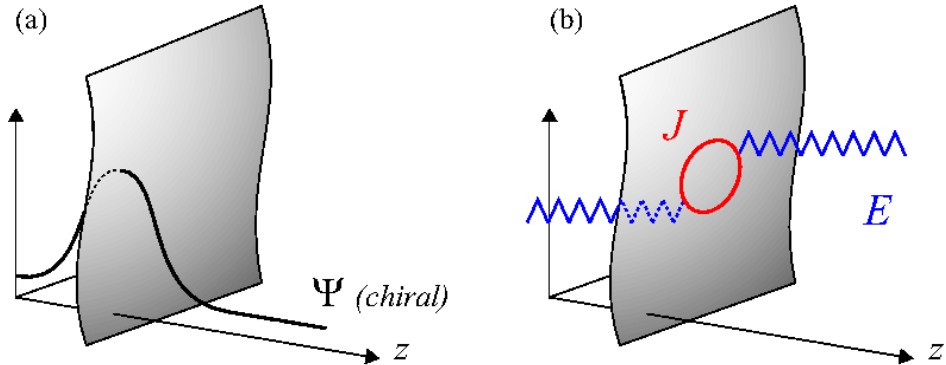}
\caption{The figures sketch the confinement of fermions and gauge
fields to a brane: (a) The wave functions of zero mode fermions are
generally found to be localised around domain walls. (b) Gauge
fields and other bosonic fields are quasilocalised via loop
corrections produced by the zero mode fermions.} \label{ghuni}
\end{center}
\end{figure}

To confine fermions to a brane, a mechanism proposed by Rubakov and
Shaposhnikov \cite{fermions1}, is considered. This mechanism is based purely on field
theoretical considerations. In their proposal, the wave functions of
fermion zero modes concentrate near existing domain walls,
generating 4-D massless chiral fermions attached to them (see Fig. \ref{ghuni}.a).
As for the gauge fields, a quasilocalisation mechanism proposed 
by Dvali, Gabadadze and Shifman \cite{gauge} is taken into account
(see refs. \cite{alt1, alt2} for alternative mechanisms).
Here, one-loop corrections produced by the interactions between
bulk gauge fields and the ``already'' confined fermions, induce gauge kinetic terms
on the brane (see Fig. \ref{ghuni}.b). The result is a 4-D effective theory consisting of
gauge fields mediating interactions between the confined fermions on
the brane.

The key ingredient of the present proposal is that the positions at
which 5-D fermions end up localised depend on their $SU(3)$ charges.
This allows, for example, to break the $\mathbf{10}$ and $\bar
\mathbf{6}$ representations of $SU(3)$ down to the lepton and quark
representations of $SU(2) \times U(1)$, respectively, and confine
them to a single brane. In this construction it is possible to
identify the Higgs with the fifth component of the localised
gauge field. Additionally, new fields inevitably appear in the
resulting 4-D effective theory. These are: a left-handed neutrino
with zero-hypercharge, and a massive vector field coupling together
the new neutrino to other left-handed leptons.

This article is organised as follows: Section \ref{S2}
discusses how the confinement of 5-D fermions is produced in the
presence of domain walls. Here, a simple toy model consisting of a
spin-1/2 fermion coupled to a scalar field through a Yukawa interaction
is presented. It is shown that if the scalar field acquires a
nontrivial vacuum solution, such as the kink profile, then a zero
mode chiral fermion confines about the domain wall. In
Section \ref{S3} this discussion is extended to include the
quasilocalisation of a $U(1)$ gauge field. Then, in Section
\ref{S4}, a more complex and interesting setup is introduced; this
time, the  bulk fields belong to $SU(3)$ representations.
It is shown that the $SU(3)$ gauge symmetry can be broken down by the
nontrivial vacuum solutions of charged scalar fields. As a result, depending on
the group representation to which bulk fermions belong, several branes
form containing different type of matter fields confined to them.
In Section \ref{S5}, we show that the fermions
confined in this way have in fact the opportunity to span a representation
of $SU(2)_{L} \times U(1)$. This allows the construction of a brane
containing leptons and quarks.
This analysis is completed in Section \ref{S6}, where the
quasilocalisation of $SU(2) \times U(1)$ gauge fields
is considered. Finally, in Section \ref{S7}, I compare the model obtained
in this way with the electroweak sector of the standard model. There I
address the fermion hierarchy problem and discuss some relevant 
phenomenological aspects of the present proposal.

\section{Confinement of fermions} \label{S2}

\begin{figure}[t] 
\begin{center}
\includegraphics[width=0.45\textwidth]{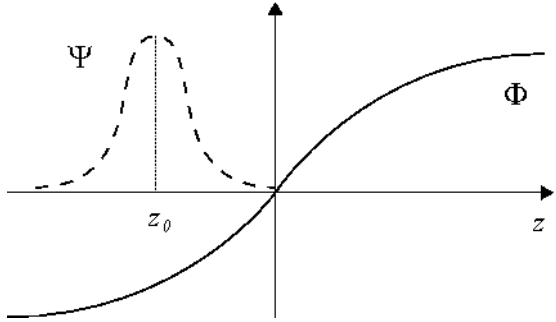}
\caption{The figure sketches the confinement of the bulk fermion
near the domain wall located at $z = 0$. The fermion wave function
is centred at position $z_{0} = - m \Delta^{2}$.} \label{shift}
\end{center}
\end{figure}

Consider a 5-D system consisting of a spin-1/2 fermion $\Psi$ and a
real scalar field $\Phi$. To describe the 5-D spacetime we use
coordinates $x^{A}$ with $A = 1, \ldots , 5$. The Lagrangian of the
system is
\begin{eqnarray}
\mathcal{L}^{(5)} = - \bar \Psi \left[\gamma^{A} \partial_{A}  + m +
y \, \Phi \right] \Psi  - \frac{1}{2} (\partial_{A} \Phi)^{2} -
V(\Phi).  \label{eq2: L-split}
\end{eqnarray}
Here $m$ is the mass of the bulk fermion $\Psi$ and $y$ is a Yukawa
coupling. Additionally, $\gamma^{A}$ are the 5-D gamma-matrices in a
basis where
\begin{eqnarray}
\gamma^{5} = \left( \begin{array}{cc} 1 & 0 \\ 0 & -1 \end{array}
\right),
\end{eqnarray}
which is the usual four-dimensional $\gamma^{5}$ matrix. For the
time being let us disregard the presence of gauge fields.
Consider also the following potential for the scalar $\Phi$
\begin{eqnarray}
V(\Phi) = \frac{\sigma}{4} \left[ \Phi^{2} - v^{2} \right]^{2}.
\end{eqnarray}
To discuss solutions to this system it is appropriate to use $z =
x^{5}$ to distinguish the extra-dimension and coordinates $x^{\mu}$,
with $\mu = 0, \cdots , 3$, to parameterise the usual 4-D spacetime.
As well known, the system admits a kink solution for the scalar field
$\Phi$, of  the form $\Phi(z) = v \tanh \left( k z \right)$,
where $k = v \sqrt{\sigma / 2} \,$. The corresponding domain wall,
centred at $z=0$, is coupled to the fermion field through the $y$-term.
Then, the equation of motion for $\Psi$ reads
\begin{eqnarray}
\left[ \gamma^{\mu} \partial_{\mu} + \gamma^{5} \partial_{z} + m +
y \, \Phi (z) \right] \Psi = 0. \label{eq: Fermion-Wall}
\end{eqnarray}
Notice that the translational invariance along $z$ is broken. It is
therefore useful to define left and right handed helicities
$\Psi_{L}$ and $\Psi_{R}$, by $\gamma^{5} \Psi_{L} = +\Psi_{L}$
and $\gamma^{5} \Psi_{R} = -\Psi_{R}$. With this convention in mind,
Eq. (\ref{eq: Fermion-Wall}) has two zero mode solutions of the form
\begin{eqnarray}
\Psi_{L,R} = A \exp \Big\{ \mp \! \int^{z}_{0} \!\!\! \left[ m + y\,
\Phi (z) \right] \, dz \Big\} \, \psi_{L,R}(x), \label{eq: solution}
\end{eqnarray}
where $\mp$ stands for left and right handed helicities, respectively,
and $\psi_{L}(x)$ and $\psi_{R}(x)$ are 4-D Weyl fermions with opposite
chiralities. Additionally, the factor $A$ is a normalisation constant
introduced in such a way that
\begin{eqnarray}
\int  d z \, |\Psi|^{2} = |\psi(x)|^{2}.
\end{eqnarray}
Thus, only one of these two solutions is normalisable: if
$y>0$ ($y<0$) then the left (right) handed fermion is normalisable.
At this stage, it is convenient to define the following ``confinement''
length scale:
\begin{eqnarray}
\Delta = \frac{1}{\sqrt{|y v k|}}.
\end{eqnarray}
Then, in general, nonzero modes solutions to Eq. (\ref{eq: Fermion-Wall})
provide a tower of states with masses of order $\Delta^{-1}$.
From now on I assume that $\Delta$ is sufficiently small so that nonzero modes
can be integrated out without affecting the theory at low energies.
Additionally, observe that if $m=0$ then the fermion wave function
is centred at $z = 0$, otherwise its localisation is shifted with
respect to the brane. To understand better this, let us analyse the linear
behaviour $\Phi \simeq v k z$ near $z = 0$ for the case $y>0$. Then,
assuming that $m^{-1} \gg k \Delta^{2}$ (so the linear expansion
$\Phi \simeq v k z$ makes sense), we obtain
\begin{eqnarray}
\Psi_{L} \sim \frac{1}{\sqrt{\Delta}}  \exp \left[ - \frac{1}{2} \Delta^{-2} (z -
z_{0})^{2} \right] \, \psi_{L}(x),
\end{eqnarray}
where $z_{0} = - m \Delta^{2}$. Thus, the fermion wave function has
a width $\Delta$ and is centred at $z_{0}$ (see Fig. \ref{shift}).
It should be noticed that the width $\Delta$ of the fermion wall is not
necessarily related to the width of the domain wall produced by the
scalar, which, here, is of order $k^{-1}$. We can now compute the
4-D effective Lagrangian for $\psi_{L}(x)$ by integrating out the
extra-dimension:
\begin{eqnarray}
\mathcal{L}^{(4)} = - \bar \psi_{L} ( \gamma^{\mu}
\partial_{\mu} ) \psi_{L}. \label{eq2: L-split-eff}
\end{eqnarray}
Notice that in the limit $\Delta \rightarrow 0$ ($z_{0} \rightarrow
0$), we obtain a thin brane characterised by
\begin{eqnarray}
\mathcal{L}^{(5)} = \delta (z) \mathcal{L}^{(4)}.
\end{eqnarray}

There is an interesting consequence related to the shift of the
fermion's positions with respect to the domain wall: Suppose a
scenario in which two bulk fermions $\Psi^{1}$ and $\Psi^{2}$, with
masses $m_{1}$ and $m_{2}$, are coupled to a wall in such a way that
$y_{1} = y >0$ and $y_{2} = -y < 0$. If in the original 5-D
Lagrangian there is a term such as $H \bar \Psi^{1} \Psi^{2} + \mathrm{h.c.}$,
where $H$ is a given bulk field (a scalar, for example), then the
4-D effective Lagrangian will contain a Yukawa term of the form
\begin{eqnarray}
\sim  ( H \, \bar \psi^{1}_{L} \, \psi^{2}_{R} + \mathrm{h.c.} )
 \, e^{ -  r^{2} / 4 \Delta^{2} }, \label{eq: Yukawa-supp}
\end{eqnarray}
where $r = r_{1} - r_{2}$ is the separation between both fermion
wave functions with $r_{1} = - m_{1} \Delta^{2}$ and $r_{2} = +
m_{2} \Delta^{2}$. This means a mechanism to exponentially suppress
4-D Yukawa couplings [in this case, for the pair ($\psi^{1}_{L}$,
$\psi^{2}_{R}$)]. This is the basis for the split fermion scenario
\cite{fermions2, fermions3, fermions4}. I will come back to this
mechanism in Section \ref{S7}, where the fermion hierarchy problem
is addressed.

\section{Quasilocalisation of gauge fields} \label{S3}

\begin{figure}[h] 
\begin{center}
\includegraphics[width=0.43\textwidth]{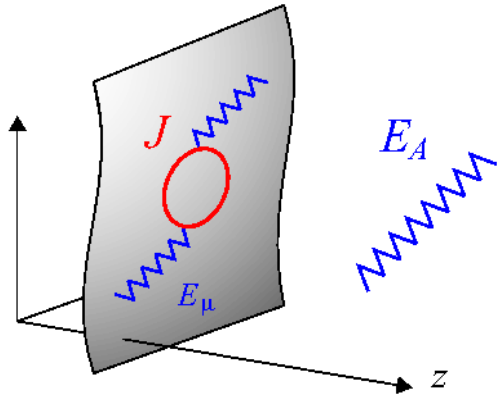}
\caption{The figure represents the propagation of gauge fields in
the bulk and the brane. If the field's wavelength is smaller than
$r_{c}$, its propagation is dominated by loop corrections taking place 
on the brane, and becomes well localised.} \label{Fquasi}
\end{center}
\end{figure}

Gauge fields can be found localised to a brane with the help
of the already confined fermions.
In the case of the quasilocalisation
of gauge fields \cite{gauge}, the interaction between the fermionic
currents at the branes and the 5-D gauge fields induces a 4-D effective 
theory. This is produced by one-loop contributions to
the effective action on the brane (see ref. \cite{DGP} for the case of
gravity). Consider, for instance,
the same setup as before but now in the presence
of a $U(1)$ gauge field $E_{A}$ with a gauge coupling $g$.
This time the Lagrangian reads
\begin{eqnarray}
\mathcal{L}^{(5)} = - \bar \Psi \gamma^{A} (\partial_{A} - i E_{A} ) \Psi
- \frac{1}{4 g^{2}} F_{A B} F^{A B} + \mathcal{L}(\Phi),
\end{eqnarray}
where $\mathcal{L}(\Phi)$ is the Lagrangian for the scalar field $\Phi$,
and $F_{A B} = \partial_{A} E_{B} - \partial_{B} E_{A}$ is the
gauge antisymmetric tensor. Then, after the fermion
confines to the wall produced by $\Phi$, in the thin wall approximation
$\Delta \rightarrow 0$, the gauge sector appears having 
the following Lagrangian
\begin{eqnarray}
\mathcal{L}^{(5)}_{\mathrm{G}} = -\frac{1}{4 g^{2}} F_{A B}
F^{A B} + \delta(z) E_{\mu} J^{\mu}(x)
\label{eq: gauge}.
\end{eqnarray}
The current term $J^{\mu}(x) = i \bar \psi_{L} \gamma^{\mu}
\psi_{L}$, localised to the brane, appears as a consequence of the
covariant derivative $D_A \Psi = (\partial_A - i E_A) \Psi$. (The
scalar component $E_{5}$ plays no role here because it couples
$\psi_{L}$ to the non-normalisable piece $\Psi_{R}$). Then, a
one-loop correction induces the following Lagrangian for $E_{\mu}$
on the brane
\begin{eqnarray}
\mathcal{L}^{(4)} = -\frac{1}{4 \lambda^{2}} F_{\mu \nu} F^{\mu \nu},
\qquad \lambda^{-2} = \frac{N}{12 \pi^{2}} \ln (\Lambda/\mu),
\end{eqnarray}
where $F_{\mu \nu} = \partial_{\mu} E_{\nu} - \partial_{\nu}
E_{\mu}$ is the antisymmetric tensor for the 4-D sector
$E_{\mu}$. Here, $\Lambda$
and $\mu$ are the ultraviolet and infrared cut-offs scales of the
5-D theory and $N$ is the number of such fermions $\psi_{L}$
participating in the loops. As a consequence, the five-dimensional 
theory describing the gauge sector near the brane, is found to have 
the form
\begin{eqnarray}
\mathcal{L}^{(5)} = -\frac{1}{4 g^{2}} F_{A B} F^{A B}
- \delta(z) \frac{1}{4 \lambda^{2}} F_{\mu \nu} F^{\mu \nu}.
\end{eqnarray}
Notice that $g^{2}$ is a dimensionful quantity, whereas $\lambda$ is
dimensionless. In fact, if the propagation of gauge bosons in this
theory is studied, one finds that there is a crossover scale $r_{c}
= g^{2}/2 \lambda^{2}$. At distances bellow this scale, the fields
propagate along the brane, whereas at larger distances, the
gauge bosons start spreading out of the brane (see Fig. \ref{Fquasi}).

\section{Confining $SU(3)$ fermions} \label{S4}

\begin{figure}[ht] 
\begin{center}
\includegraphics[width=0.85\textwidth]{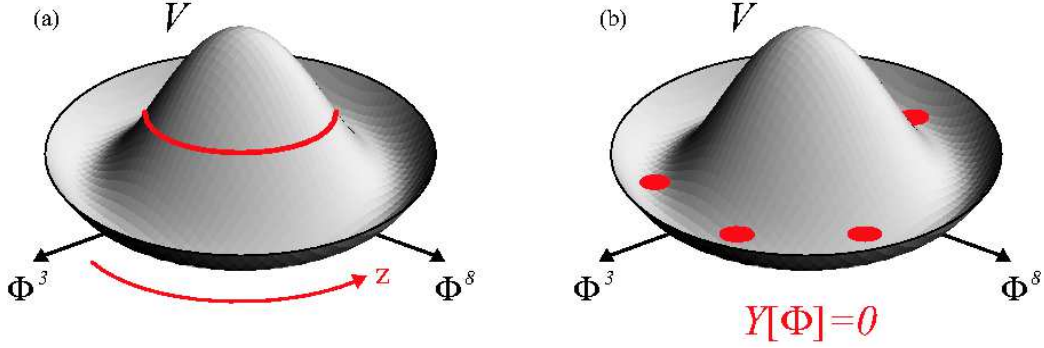}
\caption{The potential $V(\Phi)$ in terms of the scalar
components $\Phi^{3}$ and $\Phi^{8}$. (a) The loop around the hill
describes the single winding-number solution. Observe that it is
possible to parameterise the coordinate $z$ in terms of
$\Phi$-values. (b) The positions at which chiral fermions end up
confined can be deduced from the condition $Y[\Phi]=0$.}
\label{pott2}
\end{center}
\end{figure}

\begin{figure}[b] 
\begin{center}
\includegraphics[width=0.4\textwidth]{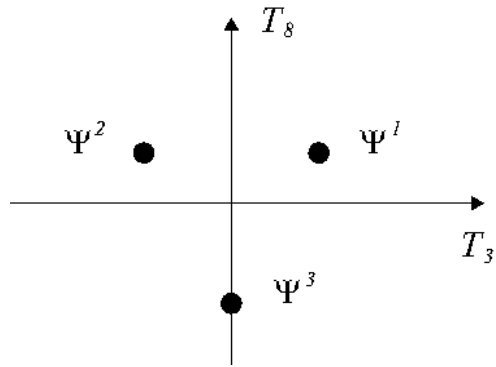}
\caption{The figure shows the $SU(3)$ charges $Q=(T_{3},T_{8})$
of fermions $\Psi^{i}$ (with $i = 1,2,3$) in the fundamental
representation $\mathbf{3}$.} \label{F2}
\end{center}
\end{figure}

\begin{figure}[ht] 
\begin{center}
\includegraphics[width=0.5\textwidth]{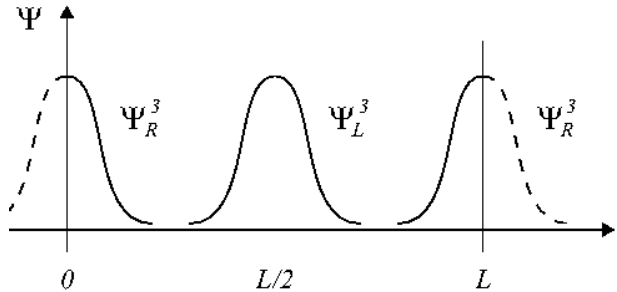}
\caption{The figure shows the way in which $\Psi^{3}$ is confined.
Same representations but with different chiralities end up in branes
located at diametrically opposite positions in the $S^{1}$ circle.}
\label{F3}
\end{center}
\end{figure}

The toy model presented in Section \ref{S2} provided a useful description of
how fermions can end up localised around a domain wall produced by
the non trivial vacuum solution of a scalar field. It is possible to
find more interesting phenomena by including a non-Abelian gauge symmetry. For
example, if the fields generating the walls are gauge-charged, then
the fermions confined to these walls have the chance of inheriting
interesting properties coming from the original symmetry. In this section we
analyse a simple case involving a five-dimensional $SU(3)$ gauge theory. 
To start, assume that spacetime is described by a 5-D manifold $M$ with topology
\begin{eqnarray}
M = \mathbb{R}^{4} \times S^{1}, \label{eq: topology}
\end{eqnarray}
where $S^{1}$ is the one-dimensional circle and $\mathbb{R}^{4}$ is
the 4-D Lorentzian space. In this case, the coordinate $z = x^{5}
\in [0,L]$ is the spatial coordinate parameterising $S^{1}$ with $L$
the size of the compact extra-dimension. Let us consider the
existence of 5-D bulk fermions transforming nontrivially under
$SU(3)$ gauge symmetry. They are described by the following
Lagrangian
\begin{eqnarray}
\mathcal{L}_{\Psi}^{(5)} = - \bar \Psi [\gamma^{A} D_{A}  + Y(\Phi) ]
\Psi. \label{eq2: Lagrangian1}
\end{eqnarray}
The covariant derivative is $D_{A} \Psi = (\partial_{A} - i
E^{\alpha}_{A} T_{\alpha}) \Psi$, where $E^{\alpha}_{A}$ are $SU(3)$
bulk gauge fields. Here $\alpha = 1, \ldots , 8$ and $T_{\alpha}$
are the $SU(3)$ generators acting on $\Psi$. Observe that there
is a coupling term $Y(\Phi)$ where $\Phi = \Phi^{\alpha}
T_{\alpha}$ is a scalar field that transforms in the adjoint
representation of $SU(3)$. In order to construct
$SU(3)$-representations we can proceed conventionally: We choose $T_{3}$
and $T_{8}$ as the Cartan generators and construct states to be
eigenvalues with charges $Q = (T_{3}, T_{8})$.
To continue, assume that $\Phi$ is dominated by the following $SU(3)$ gauge
invariant potential
\begin{eqnarray}
V(\Phi) = \frac{\sigma}{4} \left[ \Phi^{\alpha} \Phi_{\alpha} -
v^{2} \right]^{2}.
\end{eqnarray}
Nonzero vacuum expectation solutions $\langle \Phi \rangle$ are
expected and, in general, they correspond to linear combinations of
$\langle \Phi^{3} \rangle$ and $\langle \Phi^{8} \rangle$.
Furthermore, since the compact topology (\ref{eq:
topology}) has been assumed, the system admits nontrivial topological solutions.
Take for instance the case of a single winding-number solution
\begin{eqnarray}
\langle \Phi (z) \rangle = \Phi_{0} \left[ \cos (k z) T_{3} +
\sin(k z) T_{8} \right], \label{eq: Phi(z)}
\end{eqnarray}
where $k = 2 \pi / L$ and $\Phi_{0}^{2} = v^{2} - k^{2}/\sigma$.
Notice that $\langle \Phi^{8} \rangle = 0$ at $z =
0$. This solution allows to parameterise the extra-dimension in
terms of $\Phi^{3}$ and $\Phi^{8}$ values (see Fig. \ref{pott2}.a).
We can now proceed in the same way as in Section \ref{S2}:
By defining left and right handed fermions $\Psi_{L}$ and $\Psi_{R}$,
it is possible to find zero mode solutions of the form
\begin{eqnarray}
\Psi_{L,R} = A \exp \left\{ \mp \int^{z}_{0} \!\!\! Y (z) \, dz
\right\} \, \psi_{L,R}(x), \label{eq: solution2}
\end{eqnarray}
where $Y(z) \equiv Y[ \langle \Phi(z) \rangle ]$. In general,
given a nonzero v.e.v for a scalar field $\Phi (z)$, the
position $z$ at which the fermion wave function $\Psi$ is centred is
determined by the condition $Y(z) \, \Psi = 0$
(See Fig. \ref{pott2}.b). The chirality of such a state is
determined by the sign of the derivative $\partial_{z} Y(\Phi)$ at
the given position. To be more precise, if $\partial_{z} Y(\Phi) >
0$ ($\partial_{z} Y(\Phi) < 0$), then the confined fermion is left
(right) handed.
To discuss the
consequences of solution (\ref{eq: Phi(z)}) with some transparency,
let us have a look at the following simple example: Consider a Yukawa
coupling of the form
\begin{eqnarray}
Y(\Phi) = y \Phi =  y \Phi^{\alpha} T_{\alpha}, \label{eq: Yukawa}
\end{eqnarray}
and matter fields $\Psi$ belonging to the $\mathbf{3}$, the
fundamental representation of $SU(3)$. In this case the confinement
length scale must be defined as $\Delta = |y \Phi_{0} k|^{-1/2}$.
Thus again, the masses of nonzero mode states are found to be of 
order $\Delta^{-1}$.
To work out the consequences of the Yukawa coupling (\ref{eq:
Yukawa}) on the $\mathbf{3}$, it is convenient to choose 
$\Psi^{i}$ (with $i = 1,2,3$)
to have the following $SU(3)$ charges (see Fig. \ref{F2}):
\begin{eqnarray}
Q (\Psi^{1}) = (-1/2,+\sqrt{3}/6), \qquad Q(\Psi^{2}) = (+1/2,+\sqrt{3}/6),
\qquad Q(\Psi^{3}) = (0,- \sqrt{3}/3).
\end{eqnarray}
In this way, replacing (\ref{eq: Yukawa}) into (\ref{eq:
solution2}), it is possible to see that the positions at which the
fermion wave functions end up centred depend on their
$SU(3)$ charges and their chiralities. Observe, for instance, that in
the present realisation left and right-handed fermions are localised
to diametrically opposite positions in the $S^{1}$ circle.
The respective positions are: $z=0$ for $\Psi^{3}_{R}$, $z= L/2$ for
$\Psi^{3}_{L}$, $z=2L/3$ for  $\Psi^{1}_{R}$, $z=L/6$ for $\Psi^{1}_{L}$,
$z=5L/6$ for $\Psi^{2}_{R}$, and $z=L/3$ for $\Psi^{2}_{L}$.
Also, it can be seen that if $|y \Phi_{0}| \gg k$, then the widths of
wave functions become of order $\Delta$ and the overlap
between fermions located at different positions becomes very small.
Thus, the fundamental representation has been broken down to
several branes. Figure \ref{F3} shows the way in which $\Psi^{3}$ of
the fundamental representation is split.

Now it is possible to compute the 4-D effective theory for the matter fields
localised at any desired brane of our example. Let us compute, for
instance, the effective Lagrangian $\mathcal{L}_{\mathrm{eff}}$ at
the first brane ($z=0$) taking into account the presence of the
gauge field $E^{\alpha}_{A}$. In the limit $\Delta \rightarrow 0$
with $L$ fixed, the following expression is obtained
\begin{eqnarray}
\mathcal{L}_{\mathrm{eff}} = - \delta(z) \bar \psi^{3}_{R}
\gamma^{\mu} \bigg[
\partial_{\mu} + i \frac{\sqrt{3}}{3} E^{8}_{\mu} \bigg]
\psi^{3}_{R}. \label{eq: eff}
\end{eqnarray}
Here the delta function appears in the limit $\Delta \rightarrow 0$
after considering the right normalisation factor $A$ in Eq.
(\ref{eq: solution2}). Notice the appearance of the induced current
\begin{eqnarray}
J_{8}^{\mu} = - i \frac{\sqrt{3}}{3} \bar \psi^{3}_{R}
\gamma^{\mu} \psi^{3}_{R},
\end{eqnarray}
which couples to the gauge field component $E^{8}_{\mu}$ in
(\ref{eq: eff}). Then, as discussed in Section \ref{S3}, a one-loop correction
induces the following Lagrangian for $E^{8}_{\mu}$ at the brane
\begin{eqnarray}
\mathcal{L}^{(4)} = -\frac{1}{4 \lambda^{2}} (\partial_{\mu}
E^{8}_{\nu} - \partial_{\nu} E^{8}_{\mu})^{2}, \qquad
\lambda^{-2} = \frac{N}{12 \pi^{2}} \ln (\Lambda/\mu).
\end{eqnarray}
Here  $N = 1/3$, which comes from the coefficient $\sqrt{3}/3$ in
$J_{8}^{\mu}$. The final theory confined to $z=0$ corresponds to a
$U(1)$ gauge theory.


\section{Confining the electroweak model to a brane} \label{S5}

We now turn to the confinement of the electroweak model. Our
approach consists of adding a new scalar field into the model so as
to allow a richer structure to the localisation mechanism generated
by the $Y$-coupling. Then we show that leptons can be obtained from
the $\mathbf{10}$-representation of $SU(3)$, while quarks can be
obtained from the $\mathbf{\bar 6}$.
To start, assume the existence of the same scalar field $\Phi =
\Phi^{\alpha} T_{\alpha}$ (as discussed previously) and an
additional scalar field $\Theta = \Theta^{\alpha} T_{\alpha}$ also
transforming in the adjoint representation of $SU(3)$. The dynamics of
this scalar is dominated by the following $SU(3)$ gauge invariant potential
\begin{eqnarray} \label{U pot}
U \propto \left[ \Theta^{\alpha} \Theta_{\alpha} - u^{2}
\right]^{2},
\end{eqnarray}
where $u$ is a constant parameter of the theory. Now, consider the
following $Y$-coupling:
\begin{eqnarray}
Y = - y  \left( \frac{1}{2} \{ \Phi , \Theta \} - \frac{1}{4}
\Theta^{\alpha} \Phi_{\alpha} + p \frac{\sqrt{3}}{2} u \, \Phi
\right) , \label{eq: Y}
\end{eqnarray}
where $\{ \, , \}$ denotes anticommutation. In the previous
equation, $p$ is a parameter of the model that depends on the
representation on which $Y$ is acting; in the present construction 
$p = 1$ if $Y$ couples to the $\mathbf{10}$, and
$p=-1/3$ if $Y$ couples to the $\mathbf{\bar{6}}$. Other gauge
invariant terms can also be included in (\ref{eq: Y}) without
modifying the main results of this section (I will come back to this 
in Section \ref{S7}).

Let us focus on the case in which $\Theta$ acquires the following
vacuum expectation value (v.e.v.) $\langle \Theta \rangle = u \, T_{8}$.
This particular value could be due, for example, to a small
contribution $\propto \Tr \Theta^{4}$ to the potential $U$ of Eq.
(\ref{U pot}). Then, after the scalars have acquired their
respective v.e.v.'s we are left with the following $z$-dependent
coupling
\begin{eqnarray}
(y \Phi_{0} u)^{-1} Y = - \left[ ( T_{8} + p \sqrt{3}/2 )T_{8} - 1/4
\right] \sin(k z) -  \left[ T_{8} + p \sqrt{3}/2 \right] T_{3}
\cos(kz). \label{eq: Y(z)gen}
\end{eqnarray}
Similar to our previous example, in this case the widths of the
fermion wave functions become of order $\Delta$, the confining
length scale, which now is found to be
$\Delta = |y \Phi_{0} u k|^{-1/2}$.
In what follows I analyse separately the confinement of leptons
(from the $\mathbf{10}$) and quarks (from the $\mathbf{\bar 6}$).

\subsection{Leptons}

\begin{figure}[b] 
\begin{center}
\includegraphics[width=0.8\textwidth]{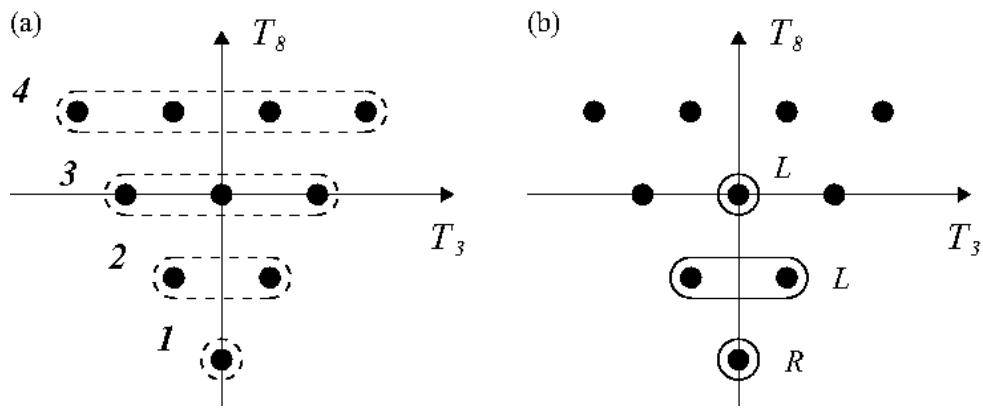}
\caption{(a) The $\mathbf{10}$ representation and its decomposition
into $SU(2)$ subgroups: this is $\mathbf{10} = \mathbf{1} \oplus
\mathbf{2} \oplus \mathbf{3} \oplus \mathbf{4}$ with charges $T_{8}
= -\sqrt{3}, -\sqrt{3}/2, \, 0, +\sqrt{3}/2$, respectively. (b)
States of the $\mathbf{10}$ that end up localised to $z=0$. The
labels $L$ and $R$ indicate the chirality of the confined states.}
\label{F4}
\end{center}
\end{figure}

Here I study the action of $Y$ on the $\mathbf{10}$ (where $p=1$)
and show that the confined fermions to the domain wall can be
identified with the usual leptons of the electroweak model.
To proceed it is convenient to consider the decomposition of $SU(3)$
into $SU(2)$ subgroups (see Fig. \ref{F4}.a).
The $\mathbf{10}$ has
the following decomposition: $\mathbf{10} = \mathbf{1} \oplus
\mathbf{2} \oplus \mathbf{3} \oplus \mathbf{4}$, with the following
$T_{8}$-charges: $T_{8} = -\sqrt{3}, -\sqrt{3}/2, \, 0,
+\sqrt{3}/2$. Using this notation, the localisation produced by the 
$Y$-coupling to the first brane at $z=0$ can be worked out. First,
observe from Eq. (\ref{eq: Y(z)gen}) that all of those states in the
$\mathbf{10}$ with $(T_{8} + \sqrt{3}/2) T_{3} = 0$ result in $Y=0$ at $z
= 0$. Then, following the reasoning of Section \ref{S4}, a chiral
fermion from each one of these states confine to $z = 0$. The
precise chirality of each state depends on the sign of $\partial_{z}
Y(z)$. In the present case, assuming $y > 0$, the confined states
are: The right-handed $SU(2)$-singlet $R \equiv
\psi^{\mathbf{1}}_{R}$ with charge $Q=(0,-\sqrt{3})$; the two
left-handed components of the $SU(2)$-doublet $L \equiv
\psi^{\mathbf{2}}_{L}$ with charges $Q=(-1/2,-\sqrt{3}/2)$ and
$Q=(+1/2,-\sqrt{3}/2)$; and only one left-handed component from the
triplet $N \equiv \psi^{\mathbf{3}}_{L}$, with charge $Q=(0,0)$.
States with opposite chirality are confined to a ``mirror-brane''
located at $z=L/2$, and any other state confines elsewhere.
Figure \ref{F4}.b shows those components of the $\mathbf{10}$ that
confine to $z=0$.


\subsection{Quarks}

\begin{figure}[t] 
\begin{center}
\includegraphics[width=0.8\textwidth]{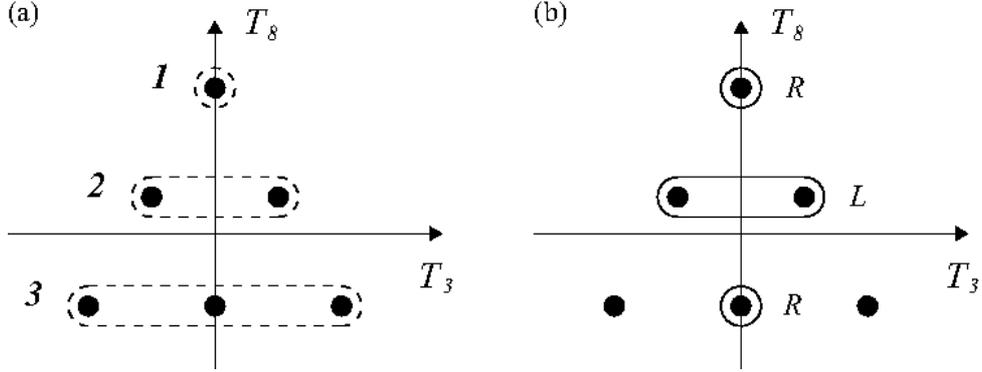}
\caption{(a) The $\mathbf{\bar 6}$ representation and its
decomposition into $SU(2)$ subgroups: this is $\mathbf{\bar 6} =
\mathbf{1} \oplus \mathbf{2} \oplus \mathbf{3}$ with charges $T_{8}
= +2 \sqrt{3}/3, + \sqrt{3}/6, -\sqrt{3}/3$ respectively. (b) States
of the $\mathbf{\bar 6}$ that end up localised to $z=0$. The labels
$L$ and $R$ indicate the chirality of the confined states.}
\label{F7}
\end{center}
\end{figure}

The case for quarks is very similar. Here, the value $p = -1/3$ in the
$Y$-coupling (\ref{eq: Y(z)gen}) needs to be considered. 
Having said this, recall that the $\mathbf{\bar 6}$
can be decomposed into $\mathbf{\bar 6} = \mathbf{1} \oplus
\mathbf{2} \oplus \mathbf{3}$ with the following $T_{8}$ charges:
$T_{8} = +2 \sqrt{3}/3, + \sqrt{3}/6, -\sqrt{3}/3$ (see Fig.
\ref{F7}.a).
Then, the following four chiral states are found to be confined: 
The right-handed $SU(2)$-singlet $\psi_{R}^{\mathbf{1}}$ with charge
$Q=(0,+2/\sqrt{3})$; the two left-handed components of the
$SU(2)$-doublet $\psi_{L}^{\mathbf{2}}$ with charges
$Q=(-1/2,+1/2\sqrt{3})$ and $Q=(+1/2,+1/2\sqrt{3})$; and only one
right-handed component from the triplet $\psi_{L}^{\mathbf{3}}$,
with charge $Q=(0,-1/\sqrt{3})$. Figure \ref{F7}.b shows those
components of the $\mathbf{\bar{6}}$ that confine to $z=0$.
These states have the appropriate
quantum numbers for this sector to be identified with the quarks
of the standard model.

\subsection{Effective theory}

We can now be more explicit by computing the effective theory for
the states confined to the brane at $z=0$. Let us analyse, for 
instance, the case for leptons. To this extent, consider the following decomposition of
the five-dimensional $SU(3)$ gauge field $E^{\alpha}_{A}$:
\begin{eqnarray}
W_{\mu}^{a} &=&  E_{\mu}^{a} \qquad  \mathrm{with} \qquad a =
1,2,3, \\
V_{\mu}^{i} &=&  E_{\mu}^{i} \qquad  \mathrm{with} \qquad i =
4,5,6,7,
\\
\phi^{i} &=&  E_{5}^{i} \qquad  \mathrm{with} \qquad i = 4,5,6,7,
\\
B_{\mu} &=&  E_{\mu}^{8}.
\end{eqnarray}
In the limit $\Delta \rightarrow 0$, other components of
$E_{A}^{\alpha}$ are decoupled from the matter fields confined to
the brane (this is because these components are coupling together
spinor fields with different chiralities that necessarily end up at
different branes). In this decomposition, the only nonzero
structure constant are: $C_{a b}^{c}$, $C_{i j}^{a}$ and $C_{i j
}^{8}$ (and obvious permutation of indices). Then, in the thin wall
approximation, the 4-D effective Lagrangian for the massless leptons
at the first brane is found to be
\begin{eqnarray}
\mathcal{L}_{\mathrm{lep}}^{(4)} = -  \bar L \Big[ \gamma^{\mu}
\partial_{\mu} - i  \gamma^{\mu} W^{a}_{\mu} T_{a}
+ i \frac{\sqrt{3}}{2} \gamma^{\mu} B_{\mu}    \Big] L - \bar R
\Big[ \gamma^{\mu}
\partial_{\mu} + i \sqrt{3}  \gamma^{\mu} B_{\mu}   \Big] R
 -  \bar N \gamma^{\mu}
\partial_{\mu} N + \mathcal{L}_{\mathrm{I}}^{(4)}, \quad \label{eq: leptons}
\end{eqnarray}
where $\mathcal{L}_{\mathrm{I}}^{(4)}$ contains interaction terms
involving $\phi^{i}$ and $V^{i}_{\mu}$
\begin{eqnarray}
\mathcal{L}_{\mathrm{I}}^{(4)} =   - i \alpha \frac{\sqrt{3}}{2}
\phi^{i} \bar R \, t_{i} \, L - i \beta V^{i}_{\mu} \bar N
\gamma^{\mu} \, s_{i} \, L + \mathrm{h.c.},  \label{eq: leptons-I2}
\end{eqnarray}
with $\alpha$ and $\beta$ coefficients emerging from the overlap
between wave functions of different widths. In the present case,
$\alpha = \beta = (5)^{1/4} / \sqrt{3}$. Additionally, $t_{i}$ and
$s_{i}$ with $i = 4,5,6,7$, are $1 \times 2$ matrices acting on the
$SU(2)$ doublet $L$ given by
\begin{eqnarray}
t_{4} = s_{6} = (1, 0), \qquad t_{5} = - s_{7} = i ( 1, 0), \qquad
t_{6} = s_{4} = (0, 1), \qquad t_{7} = - s_{5} = i (0, 1).
\end{eqnarray}
These matrices appear as a consequence of the action of the operators 
$T_{i}$'s on those states of the $\mathbf{10}$ that are later
identified with $L$.

\subsection{About the other branes}

\begin{figure}[h] 
\begin{center}
\includegraphics[width=0.5\textwidth]{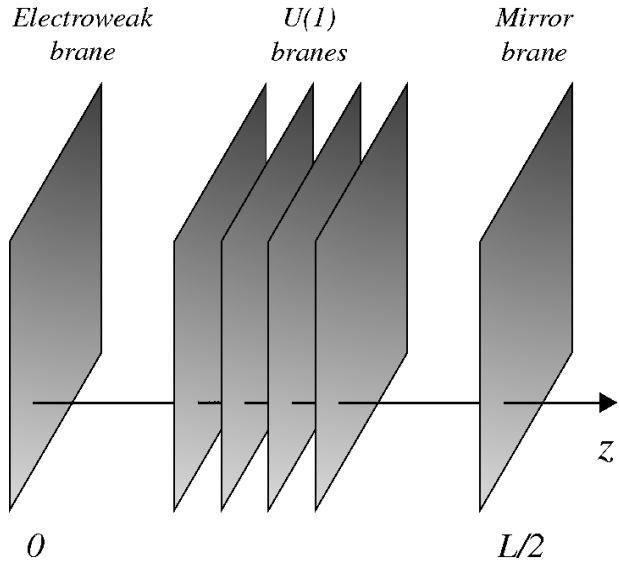}
\caption{The figure sketches the disposition of branes in the 5-D
bulk. The electroweak brane is located at $z=0$, while the mirror
brane (a copy of the first brane but containing matter with opposite
chirality) is located at $z=L/2$.}
\label{other-b}
\end{center}
\end{figure}

To finish, let us briefly mention that other branes are also formed
in the bulk. They appear from the localisation of the rest of the
states of the $\mathbf{\bar 6}$ and $\mathbf{10}$ representations.
The most interesting brane is the ``mirror brane'' at $z=L/2$, which
contains a copy of the electroweak model obtained at the first brane
$z=0$, but with states having opposite chiralities. The rest of the
branes (also determined by the condition $Y=0$) all contain
different versions of $U(1)$ Abelian gauge theories (see Fig. \ref{other-b}).

\section{Localisation of $SU(2) \times U(1)$ gauge fields} \label{S6}

The effective theory of Eq. (\ref{eq: leptons}) is invariant under
$SU(2) \times U(1)$ gauge symmetry. Before fully identifying this
theory with the electroweak model in Section \ref{S7}, it is 
important to consider the localisation of the gauge fields
$W^{a}_{\mu}$, $B_{\mu}$, $\phi^{i}$ and $V_{\mu}^{i}$ to the brane.
To start, observe that there is an $SU(2) \times U(1)$ 
current term attached to the brane of the form
\begin{eqnarray}
E^{\alpha}_{A} J^{A}_{\alpha} = W^{a}_{\mu} J_{a}^{\mu}(x) + B_{\mu}
J^{\mu}(x)  + V^{i}_{\mu} J_{i}^{\mu}(x) + \phi^{i} J_{i}(x).
\end{eqnarray}
Since the effective terms for gauge fields are induced by loop corrections
from these currents, the transformation properties of $J^{A}_{\alpha}(x)$
are transferred to the quasilocalised gauge fields. Therefore,
the 4-D induced action for $W_{\mu}^{a}$, $V_{\mu}^{i}$, $B_{\mu}$
and $\phi^{i}$ at the first brane ($z=0$) becomes
\begin{eqnarray}
\mathcal{L}^{(4)}_{\mathrm{G}} = - \frac{1}{4 \lambda^{2}_{H}}
H^{a}_{\mu \nu} H_{a}^{\mu \nu} - \frac{1}{4 \lambda^{2}_{G}} G_{\mu
\nu} G^{\mu \nu}  - \frac{1}{2 \lambda^{2}_{\phi}} |D \phi|^{2}
 - \frac{1}{4 \lambda^{2}_{Q}} Q^{i}_{\mu \nu} Q_{i}^{\mu \nu}
+ \mathcal{L}_{V} .  \label{eq: induced}
\end{eqnarray}
Here $H_{\mu \nu}^{a}$, $Q_{\mu \nu}^{i}$, $G_{\mu \nu}$ and
$D_{\mu} \phi^{i}$ are defined as
\begin{eqnarray}
H_{\mu \nu}^{a} &=&
\partial_{\mu} W_{\nu}^{a} -
\partial_{\nu} W_{\mu}^{a} +  C^{a}_{b c} W_{\mu}^{b}
W_{\nu}^{c}  ,  \nonumber\\
Q_{\mu \nu}^{i} &=& \partial_{\mu} V_{\nu}^{i} -
\partial_{\nu} V_{\mu}^{i} +  C^{i}_{a j} W_{\mu}^{a} V_{\nu}^{j}
+ C^{i}_{j a} V_{\mu}^{j} W_{\nu}^{a} + C^{i}_{8 j} B_{\mu} V_{\nu}^{j} + C^{i}_{j 8} V_{\mu}^{j}  B_{\nu} , \nonumber\\
G_{\mu \nu} &=& \partial_{\mu} B_{\nu} -
\partial_{\nu} B_{\mu}, \nonumber\\
D_{\mu} \phi^{i} &=& \partial_{\mu} \phi^{i} + C^{i}_{a j}
W^{a}_{\mu} \phi^{j} +  C^{i}_{8 j} B_{\mu} \phi^{j}  .
\end{eqnarray}
Additionally, in Eq. (\ref{eq: induced}) there is $\mathcal{L}_{V}$, 
which contains interaction terms between the
vector field $V_{\mu}^{i}$ and the rest of the induced fields
\begin{eqnarray}
\mathcal{L}_{V} =  - \frac{1}{4 \lambda^{2}_{1}} \left( R_{\mu
\nu}^{a} R^{\mu \nu}_{a} + K_{\mu \nu} K^{\mu \nu}   \right)  -
\frac{1}{2 \lambda^{2}_{2}} H^{a}_{\mu \nu} R^{\mu \nu}_{a}    -
\frac{1}{2 \lambda^{2}_{3}} G_{\mu \nu} K^{\mu \nu} - \frac{1}{2
\lambda^{2}_{4}} \left( S_{\mu}^{a} S^{\mu}_{a} + S_{\mu} S^{\mu}
\right), \label{eq: L-V}
\end{eqnarray}
where we have defined: $R_{\mu \nu}^{a} =  C^{a}_{i j} V_{\mu}^{i}
V_{\nu}^{j}$, $S_{\mu}^{a} =  C^{a}_{i j} V_{\mu}^{i} \phi^{j}$,
$S_{\mu} = C_{i j}^{8} V^{i}_{\mu} \phi^{j} / \sqrt{3}$ and $K_{\mu
\nu} =  C^{8}_{i j} V_{\mu}^{i} V_{\nu}^{j} / \sqrt{3}$. Finally,
the various couplings $\lambda_{H}$, $\lambda_{G}$, $\lambda_{Q}$
and $\lambda_{\phi}$ in (\ref{eq: induced}), and $\lambda_{1}$,
$\lambda_{2}$, $\lambda_{3}$ and $\lambda_{4}$ in (\ref{eq: L-V})
are, in general, found to be of the form
\begin{eqnarray}
\frac{1}{\lambda^{2}} = \frac{N}{12 \pi^{2}} \ln
\frac{\Lambda}{\mu}, \label{eq: lambda}
\end{eqnarray}
where $N$ measures the number of fermions present in the different
loops, taking also into account the values of the various
$SU(3)$-charges and combinatorics. For example, we have
\begin{eqnarray}
N_{H} = \mathrm{Tr} \left( T_{3}^{2} \right), \quad \mathrm{and}
\quad N_{G} = \mathrm{Tr} \left( T_{8}^{2} \right),
\end{eqnarray}
where the traces run over all charged fermions taking place in the
loops inducing the first and second terms of (\ref{eq: induced}).

\section{Discussion} \label{S7}

\begin{figure}[t] 
\begin{center}
\includegraphics[width=0.8\textwidth]{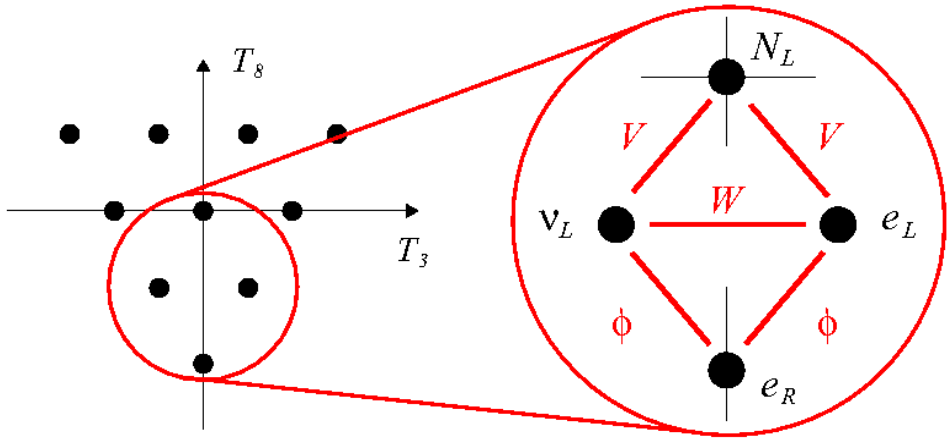}
\caption{The figure shows the bosonic fields mediating interactions 
between those states of the $\mathbf{10}$ that localise to the electroweak
brane (leptons).} \label{zoom10}
\end{center}
\end{figure}

We can now compare this theory with the lepton sector of the
electroweak model. The two left-handed components $L$ and the
right-handed fermion $R$ can be identified with the usual
counterparts of the electroweak model; namely, the pair of 
left-handed leptons $(e_{L} , \nu_{L})$ and the right handed 
electron $e_{R}$. Also, $W^{a}_{\mu}$ and $B_{\mu}$ can be 
identified with the usual $SU(2) \times U(1)$ gauge fields with couplings
$g_{1} = \lambda_{H}$ and $g_{2} = \sqrt{3} \lambda_{G}$
respectively. One of the most interesting aspects of this model,
however, is the appearance of two additional fields: A
vector field $V_{\mu}^{i}$ and the left-handed neutrino $N$ of 
zero-hypercharge. Observe that this neutrino interacts only
with the other left-handed particles $L$ through the vector field
$V_{\mu}^{i}$. Figure \ref{zoom10} shows the states of the 
$\mathbf{10}$ that confine to the electroweak brane ---leptons--- 
and the bosonic degrees mediating interactions between them.

If we further assume that $\phi^{i}$ develops a nonzero v.e.v.
$|\phi| = \phi_{0}$, then $\phi^{i}$ takes the role of the Higgs 
field. If this is the case,
two of the chiral states ($R$ and one of the $L$'s) mix together to
form a massive electron, while the other two remain massless (neutrinos).
The electroweak parameters are then found to be as follows: The
electron mass is $m_{e}^{2} = 3 \phi_{0}^{2} \lambda_{\phi}^{2} /
2$, the $W$-boson mass is $M_{W}^{2} = \phi_{0}^{2}
\lambda_{H}^{2}/4$, and the electroweak angle is $\sin^{2}
\theta_{W} = 3 \lambda_{G}^{2}/(\lambda_{H}^{2} + 3
\lambda_{G}^{2})$. Observe that the masses of 
leptons and $W$ bosons are of the same order.
What is more, the quark masses are found to be proportional to 
$\phi_{0} \lambda_{\phi}$, of the same order as the electron mass.
This is because the $\lambda$ couplings are all of the same order. 
To generate a hierarchy between fermions and gauge bosons,
as observed in nature, it is simple a matter of generalising 
$Y$ to contain other gauge invariant interactions. 
For example, let us consider a new coupling
$Y'$ of the form
\begin{eqnarray}
Y' = Y - y \, q \, v \, \Theta,
\end{eqnarray}
where $q$ is a dimensionless coefficient that could depend on the
representation on which $Y'$ is acting (observe the similarity of
the new term with the old one $- y \, p \, u \, \Phi$, in $Y$).
Then, after the scalars have acquired the v.e.v. discussed before,
the $Y'$ coupling becomes
\begin{eqnarray}
Y'(z) = Y(z) - q (y v u ) T_{8}.
\end{eqnarray}
The second term of this expression resembles the 5-D mass term of
Eq. (\ref{eq2: L-split}). Therefore, the fermion wave functions will 
split around the branes and an exponential factor, like the one
of Eq. (\ref{eq: Yukawa-supp}),  will appear suppressing the
couplings of Eq. (\ref{eq: leptons-I2}). This results in a hierarchy
between the mass scales of quarks, leptons and gauge
bosons. Observe that in the definition of $Y$ we could also include terms
proportional to $\Phi^{2}$ and $\Theta^{2}$ with coefficients
depending on the representation. They would provide additional
terms contributing to the split of fermions around the brane.
\begin{figure}[t] 
\begin{center}
\includegraphics[width=0.9\textwidth]{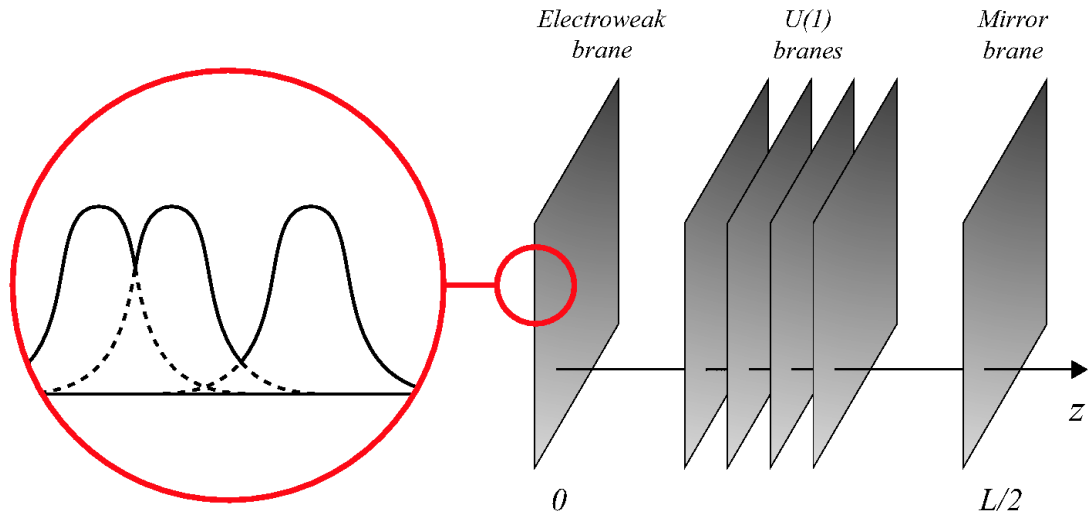}
\caption{The electroweak brane consists 
of different fermions with overlapping wave functions.}
\label{F8}
\end{center}
\end{figure}

Very important for this model is that the
existence of $V^{i}_{\mu}$ has no conflicts with observations.
Fortunately, the mechanism generating the fermion hierarchy 
is also suppressing the couplings between
$V$ and leptons. Additionally, in the case of a nonzero v.e.v. $\phi_{0}$, the
four-component vector field $V_{\mu}^{i}$ becomes massive, with
$M_{V}^{2} = \phi_{0}^{2} \lambda_{\phi}^{2} \lambda_{Q}^{2} / 4
\lambda_{4}^{2}$. In fact, the non-observation of 
$V$-bosons pair-production at LEP \cite{exp-W} is an indication 
of the constraint
\begin{eqnarray}
 M_{V} > 104 \mathrm{GeV}.
\end{eqnarray}
Nevertheless, we should not expect a value $M_{V}$ significantly 
higher than $M_{Z}$ and $M_{W}$. If this is the case, then we could expect new
phenomena associated with extra-dimensions in lepton-collider
experiments in the near future.

There are several interesting questions that can be raised about the
present model. For example, it would be important to analyse how to
include the mixing between different families of leptons and quarks.
In the case of leptons, for instance, the new neutrino $N$ could be
playing some relevant role in the mixing of neutrinos. Also, it
still remains to understand a mechanism to obtain the appropriate
potential for the Higgs field $\phi^{i}$.


\ack I would like to acknowledge support from the Swiss National 
Science Foundation. My gratitude also to the
Cambridge Philosophical Society for funding part of my trip to Corfu.

\section*{References}



\begin{thebibliography}{99}



\bibitem{Palma} Gonzalo A. Palma, Phys. Rev. D {\bf 73}, 045023 (2006).


\bibitem{Fairlie} D. B. Fairlie, Phys. Lett. B {\bf 82}, 97 (1979).


\bibitem{Manton} N.S. Manton, Nucl. Phys. B {\bf 158}, 141 (1979).



\bibitem{G-H 1} L.J. Hall, Y. Nomura and D. R. Smith, Nucl. Phys. B
{\bf 639}, 307 (2002).

\bibitem{G-H 2} G. Burdman and Y. Nomura, Nucl. Phys. B {\bf 656}, 3 (2003).

\bibitem{G-H 3} C. Csaki, C. Grojean and H. Murayama, Phys. Rev. D {\bf 67},
085012 (2003).

\bibitem{G-H 4} C. A. Scrucca, M. Serone and L. Silvestrini, Nucl. Phys. B
{\bf 669}, 128 (2003).

\bibitem{G-H 5} N. Haba and Y. Shimizu, Phys. Rev. D {\bf 67},
095001 (2003).

\bibitem{G-H 6}
G. Dvali, S. Randjbar-Daemi and R. Tabbash, Phys. Rev. D {\bf 65}, 064021 (2002).

\bibitem{G-H 7} I. Gogoladze, Y. Mimura, S. Nandi, Phys. Lett. B {\bf 560},
 204 (2003); Phys. Rev. D {\bf 69}, 075006 (2004).

\bibitem{G-H 8} K. Agashe, R. Contino and A. Pomarol, Nucl. Phys. B
{\bf 719}, 165 (2005).



\bibitem{fermions1} V. A. Rubakov and M. E. Shaposhnikov, Phys. Lett. B {\bf 125}, 136 (1983).

\bibitem{fermions2} N. Arkani-Hamed and M. Schmaltz, Phys. Rev. D {\bf 61}, 033005 (2000).

\bibitem{fermions3} N. Arkani-Hamed, Y. Grossman and M. Schmaltz, Phys. Rev. D {\bf 61}, 115004 (2000).

\bibitem{fermions4} E. A. Mirabelli and M. Schmaltz, Phys. Rev. D {\bf 61}, 113011 (2000).



\bibitem{gauge} G. Dvali, G. Gabadadze and M. Shifman, Phys. Lett. B {\bf 497}, 271 (2001).

\bibitem{alt1} G. Dvali, M. Shifman, Phys. Lett. B {\bf 396}, 64 (1997).

\bibitem{alt2} S. L. Dubovsky, V. A. Rubakov and P. G. Tinyakov, J. High Energy Phys. 08 (2000) 041.


\bibitem{DGP} G. Dvali, G. Gabadadze, M. Porrati, Phys. Lett. B {\bf 485}, 208 (2000).




\bibitem{exp-W} G. Abbiendi, {\it et al.} (The OPAL Collaboration), Eur. Phys. J. C
{\bf 26}, 321 (2003).



\end{thebibliography}
\end{document}